# π and 4π Josephson Effects Mediated by a Dirac Semimetal

W. Yu[1], W. Pan[1], D. L. Medlin[2], M. A. Rodriguez[1], S. R. Lee[1], Zhi-qiang Bao[3], and F. Zhang[3]

[1]*Sandia National Laboratories, Albuquerque, New Mexico 87185, USA*

[2]*Sandia National Laboratories, Livermore, CA 94551, USA*

[3]*Department of Physics, University of Texas at Dallas, Dallas, TX, USA*

$Cd_3As_2$ is a three-dimensional topological Dirac semimetal with connected Fermi-arc surface states. It has been suggested that topological superconductivity can be achieved in the nontrivial surface states of topological materials by utilizing the superconductor proximity effect. Here we report observations of both π and 4π periodic supercurrents in aluminum-$Cd_3As_2$-aluminum Josephson junctions. The π period is manifested by both the magnetic-field dependence of the critical supercurrent and the appearance of half-integer Shapiro steps in the a.c. Josephson effect. Our macroscopic theory suggests that the π period arises from interference between the induced bulk superconductivity and the induced Fermi arc surface superconductivity. The 4π period is manifested by the missing first Shapiro steps and is expected for topological superconductivity.

A three-dimensional topological Dirac semimetal has symmetry-protected linear energy dispersions in all three momentum directions giving rise to many unusual properties and phenomena [1-3]. One of the exciting theoretical predictions is the existence of connected Fermi-arc states on the surfaces of a Dirac semimetal [1-3]. It has been predicted that topological superconductivity [4,5] can be achieved by coupling nontrivial surface states, for example, through the so-called proximity effect to conventional superconductors [6]. Such topological superconductivity may harbor Majorana zero modes, which are potential building blocks for topological quantum computation [7]. Among many known Dirac semimetals, $Cd_3As_2$ [3] has uniquely useful properties: it is air stable, and it can be readily grown by vapor transport [8], pulsed laser deposition [9], and more recently molecular beam epitaxy [10]. Thus, $Cd_3As_2$ is a particulary promising material for topological quantum computation applications.

Since the discovery of $Cd_3As_2$ as a stable Dirac semimetal [3,11,12,8], multiple approaches have been used to realize topological superconductivity [13] in this material. In initial efforts about two years ago, two groups using point contact technique, reported tentative evidence of unconventional superconductivity in $Cd_3As_2$ [14,15]. However, in these studies, the hallmark of superconductivity, a zero-resistance state, was not observed. Later, by using the high-pressure technique, superconductivity and the zero-resistance state were achieved [16]. Yet, further x-ray diffraction (XRD) analysis showed that the pressure induced superconducting transition was accompanied by a structure transition [16,17]. Thus, it is not clear whether the induced superconductivity is topological.

On the other hand, because the surface states in $Cd_3As_2$ are similar to those of topological insulators, topological superconductivity can also be achieved in $Cd_3As_2$ through the proximity effect that couples its surface states to an s-wave Josephson junction [6,18]. The induced pairing in the surface states gives rise to topologically protected gapless Andreev bound states, i.e., Majorana zero modes, whose energies vary $4\pi$ periodically with the superconducting phase difference across the junction [19-24]. This $4\pi$ periodicity, which can be manifested in the a.c. Josephson effect via missing odd integer Shapiro steps, has been regarded as a compelling experimental signature of realizing topological superconductivity [20-24]. In addition, the induced pairing in the coexisting bulk states can lead to normal superconductivity, with a $2\pi$ period Josephson effect. The interference between the surface topological superconductor and bulk normal superconductor can further produce a $\pi$ period Josephson effect, which can be revealed by the appearance of half-integer steps in the a.c. Josephson effect [see the Supplemental Material (SM) for details].

In this Letter, we report the observation of superconducting supercurrent states in aluminum-$Cd_3As_2$-aluminum Josephson junctions. The most important outcome of these experiments is that both $\pi$ and $4\pi$ periodic supercurrent states are clearly observed. Complementary theoretical modeling by a resistively shunted junction model [25-27] suggests that the $\pi$ period is due to interference of the induced bulk superconductivity and the induced Fermi-arc surface superconductivity, whereas the $4\pi$ period is expected in a topological superconductor [19-23]. Taken together, our results provide compelling evidence that we have achieved Fermi-arc topological superconductivity in the Dirac semimetal $Cd_3As_2$.

The $Cd_3As_2$ polycrystalline ingots used in this study were purchased from a commercial source. Our detailed XRD analysis (Fig. S1 in the SM) shows that the ingots only contain a single-phase of $Cd_3As_2$. The extracted lattice constants are a = 1.2680(2) nm and c = 2.5308(6) nm, consistent with both recent reports [28-30,17] and ICDD powder-diffraction-file data [31]. We note that reported lattice constants vary slightly with each group suggesting that the various forms of $Cd_3As_2$ currently in use (polycrystalline powders, polycrystalline ingots, and single crystals) have subtly different residual-strain states, stoichiometries, or defect contents. In the Supplemental Material (SM) [25], we also show the 2D pole figures for a $Cd_3As_2$ ingot (Fig. S2) sampled from the batch used in our transport studies. As discussed in the SM, the pole-figure data indicates that the $Cd_3As_2$ ingots used in the present work consist of randomly oriented, large-grain polycrystals. The large grain size and random orientation necessitated use of specialized, area-detector-based diffraction methods in order to reliably verify the $Cd_3As_2$ phase and to characterize its orientation distribution; details of these methods are separately reported by Rodriguez *et al* [32].

The mechanical exfoliation method is used to obtain $Cd_3As_2$ thin flakes from the initial ingot materials (see SM [25]). We note that deformation during the exfoliation may introduce crystal defects beyond those of the initial polycrystalline ingot. In preparing our $Cd_3As_2$ devices, we always choose the most flat and shiny flakes for sample fabrication. In Fig. S3 in the SM [25], we show the observation of quantum oscillations [33-39] in one of the thin flakes made with conventional ohmic contacts, indicating this material is of high quality and well-suited to achieve transparent Josephson junctions. To fabricate superconductor-$Cd_3As_2$-superconductor Josephson junctions, we deposit thin flakes of $Cd_3As_2$ on a $Si/SiO_2$ substrate followed by electron beam lithography to define electrical contacts. Ti/Al (10 nm / 200 nm) bilayers are used as the superconductor electrodes to ensure good electrical contact and supercurrent in the junctions at low temperatures. In Fig. S4 in the SM [25], we show a transmission electron microscopy (TEM) image of the cross section of one fabricated device. Smooth interfaces are seen between the substrate and $Cd_3As_2$ flake as well as between the $Cd_3As_2$ and contact materials, respectively. The observed thickness of $Cd_3As_2$ flake is ~ 200 nm. The inset of Figure 1a shows a scanning electron microscopy (SEM) image of a typical as-fabricated device (Junction A). In this device, the junction length $L_j$ = 90 nm and the junction area $S_j$ = ~ 4.5×10$^{-14}$ m$^2$ (note that the adjacent electrode widths are ~ 500 nm and provide a length scale for the image). Several junctions of different sizes were fabricated and studied. Consistent results were obtained in all the junction devices. In this Letter, we mainly focus on the data measured in Junction A. Additional data from Junction B are included in the SM [25].

Figure 1a shows the resistance of Junction A as a function of temperature (*R-T* trace) at zero magnetic field (*B*). The resistance is constant at high temperatures, with a sharp drop occurring at a temperature *T* ~ 1.1 K due to the onset of a superconducting transition of Ti/Al electrodes. The resistance continues to decrease as the junction cools down and reaches the zero-resistance state below the transition temperature $T_c$ = 0.8 K.

Figure 1b displays the current-voltage (*I-V*) characteristics of the junction measured at *T* = 12 mK. For large d.c. currents, $I_{dc}$, the *I-V* curve follows a straight line with a non-zero slope indicating a normal state of the junction. From the slope of the *I–V* curve, the normal-state resistance of $R_n$ ~ 23.8 Ω is extracted. In the regime $|I_{dc}|$ < 4.6 μA, the voltage across the junction is zero demonstrating a robust Josephson supercurrent state. The critical current $I_c$ ~ 4.6 μA is marked by the arrow. The $I_c R_n$ product gives a characteristic voltage of ~ 110 μV for Junction A. This value is consistent with the transition temperature of 0.8K (or an energy gap Δ ~ 120 μV, assuming Δ = 1.76$k_B T_c$), indicating that a highly transparent interface has been achieved in our junction device.

To further investigate the properties of the supercurrent states within the junction, we studied the magnetic-field dependence of the critical current $I_c$. Figure 2a shows a 2D color plot of d*V*/d*I* as a function of the d.c. bias current and the magnetic field, with the junction held at 12 mK. We note here that d$V$/d$I$ is purposely used in this plot to enhance the contrast. The uniformly colored blue area shows the superconductivity regime and the sharp edge of the region highlights the value of $I_c$. The supercurrent completely vanishes at about 40 mT, at which point Junction A becomes the normal state. Indeed, as shown in Fig. 2b, the magneto-resistance is zero around $B = 0$ T, increases quickly around 30 mT, and then saturates when $B$ is larger than 40 mT and the junction has reached the normal state. In Fig. 2c, we plot the value of the critical current as a function of magnetic field. It is clearly seen that the critical current $I_c$ displays a non-monotonic $B$ field dependence around $B = 0$. $I_c$ first increases with increasing B, reaching a maximal value of ~ 4.8 µA at $B$ ~ 5 mT, and then decreases continuously. We note that this magnetic-field-induced enhancement in $I_c$ has recently been reported in InAs material systems [40,41] and argued as a result of strong spin-orbit coupling. Because spin-orbit coupling is known to be very strong in $Cd_3As_2$ [42], we infer that the same mechanism also applies to our device.

Further, more detailed examination of Figs. 2a and 2c proves yet more revealing. First, we note that $I_c$ does not exhibit the oscillatory Fraunhofer-like pattern as expected for Josephson junctions [43]. This is due to the extremely small junction size of our device. Given the junction area $S_j$ of ~ 4.5×10$^{-14}$ m$^2$, the first minimum of $I_c$ is expected to occur at $B = \Phi_0/ S_j$ ~ 46 mT for the 2π period in a Josephson junction. Here, $\Phi_0 = h/2e = 2.067\times10^{-15}$ Wb is the magnetic flux quantum, $h$ the Plank constant, and $e$ the electron charge. This value is larger than the $B_c$ of 40 mT in our device, thus explaining the lack of oscillatory $I_c$ behavior. Second, we note that the dependence of $I_c$ on magnetic field shows a surprising kink behavior at a smaller magnetic field of ~ 22 mT. Intriguingly, $B$ ~ 22 mT corresponds to $\Phi_0/2$ or a π-period supercurrent with respect to phase difference. The appearance of this π period is corroborated below by the observation of the half-integer Shapiro steps in a.c. Josephson measurements [44]. This π period, as we will argue later, is very likely due to the interference between the bulk superconductivity and the surface conductivity in our junction device.

The a.c. Josephson effect is studied by measuring the *I-V* characteristics with microwave radiation of varying frequencies and powers. We caution here that the microwave power we quote in the manuscript is the value measured at room temperature external to our cryostat. Since microwave loss in the semi-rigid coax cable connecting to our low-temperature experimental setup is frequency dependent, the exact microwave power delivered at the device level differs from the external measurements. Figure 3a shows *I-V* curves measured at a microwave frequency of $f = 6.5$ GHz, with the junction temperature held at 200 mK. Shapiro steps are clearly visible at $V_{dc} = Nhf/2e$ (where N is an integer) that relates to the $2\pi$ period supercurrent. As the microwave frequency is increased to a higher value, for example, $f = 10.5$ GHz, extra steps start to appear at half-integers when microwave power is increased. For example, in the blue trace of Fig. 3b with a low irradiation power of $P = -3$ dBm, the Shapiro steps occur at N = 1, 2, 3… When the irradiation power increases to 2 dBm as in the green trace, fractional Shapiro steps emerge at half-integers, i.e., $V_{dc} = Nhf/4e$. Half-integer steps are also observed at other high frequencies.

Half-integer Shapiro steps have been occasionally observed before [44-49] in superconducting quantum interference devices (SQUID). In a SQUID, the supercurrent bifurcates into the two paths of the ring and interferes after the recombination. The interference pattern strongly depends on the phase difference between the two paths as well as the self-inductance of the ring. Because of the coexistence of bulk-state and surface-state channels in the Dirac semimetal, our Josephson junction can be viewed effectively as a SQUID with bulk-surface interference, as shown in Fig. 4. As such, our device can naturally be described by a resistively shunted junction model [25,26]. Following the perturbative analysis in Ref. [27], the current in shunted parallel junctions may be written (see the SM for details):

$$I = 2\text{Im}\left[xe^{i\phi_0}\sum_{k=-\infty}^{\infty}J_{-k}(a)e^{i(\omega_0-k\omega)\tau} + \pi\beta y^2 e^{i2\phi_0}\sum_{k=-\infty}^{\infty}J_{-k}(2a)e^{i(2\omega_0-k\omega)\tau}\right],$$

where $x = \cos\pi\Psi$, $y = \sin\pi\Psi$, $\omega_0 = V_0/RI_c$, $a = 2\pi V_1/\omega_f\Phi_0$, $\omega = \Phi_0\omega_f/2\pi RI_c$, $\beta = LI_c/\Phi_0$, $\tau = 2\pi RI_c t/\Phi_0$, $V = V_0 + V_1\cos(\omega_f t)$, $\phi_0$ is an initial constant, $\Phi_0$ the magnetic flux quantum, $\Psi$ the phase difference between the bulk and surface channels, and $J_k$ is the $k$-th order Bessel function. As futher discussed in the SM, this model indicates that the π-period most likely arises from a sizable self-inductance in the junction and a spontaneous phase difference (close to π) between the induced bulk and surface superconductivity that give rise to the pronounced half-integer steps. More detailed microscopic simulations are under development and will be dedicated to a separate publication.

One of the hallmarks of topological superconductivity is the existence of a 4π-period supercurrent in a Josephson junction, as a consequence of the formation of topologically protected gapless Andreev bound states whose energies vary 4π periodically with the superconducting phase differences across the junction [19-23]. This 4π period can be manifested either in the Fraunhofer pattern (which as already noted, we cannot access) or in a.c. Josephson effect via missing odd integer Shapiro steps (for which we probed further). In Figure 3c, we show *I-V* curves taken at *f* = 3 GHz. Similar to previous works [50,51], the disappearance of the first (N = ±1) Shapiro steps is observed at relatively low frequencies in our sample. At relatively high frequencies, the N = ±1 steps reappear, as shown in Fig, 3a. Also, consistent with previous works [44,45], only the N=±1 steps disappear. Other odd steps, N = ±3, 5, etc. remain visible at all of the frequencies studied in this work. The observed disappearance of the N=±1 steps at lower frequencies suggests the realization of a topological surface superconductivity.

In summary, we fabricated superconductor-$Cd_3As_2$-superconductor junctions out of $Cd_3As_2$ flakes. Transport properties of the junctions were investigated. At low temperature, we observed Josephson supercurrent in $Cd_3As_2$ that was induced by the proximity effect. The dependence of the supercurrent on the magnetic field was studied in detail and revealed multiple unusual behaviors. The critical current $I_c$ exhibited a weak minimum at $\Phi_0/2$ indicating π-period Josephson effect. An anomalous $I_c$ enhancement by the magnetic field was also observed due to strong spin-orbit coupling in $Cd_3As_2$. Finally, Shapiro steps and anomalous step behavior were observed when microwave irradiation was applied and then varied in power and frequency, demonstrating the presence of the a.c. Josephson effect within the junctions. Particularly, several half-integer steps and one missing odd-integer step were clearly evidenced. The realization of an unconventional Josephson supercurrent in $Cd_3As_2$ with surface-bulk interference presents an important step towards searching for topological surface superconductivity and non-Abelian Majorana zero modes.

The work at Sandia National Labs was supported by a Laboratory Directed Research and Development project. Device fabrication was performed at the Center for Integrated Nanotechnologies, an Office of Science User Facility operated for the U.S. Department of Energy (DOE) Office of Science. Sandia National Laboratories is a multimission laboratory managed and operated by National Technology and Engineering Solutions of Sandia, LLC., a wholly owned subsidiary of Honeywell International, Inc., for the U.S. Department of Energy's National Nuclear Security Administration under contract DE-NA-0003525. The work at UT Dallas was supported by UT Dallas research enhancement funds.

**Figures and Figure Captions**

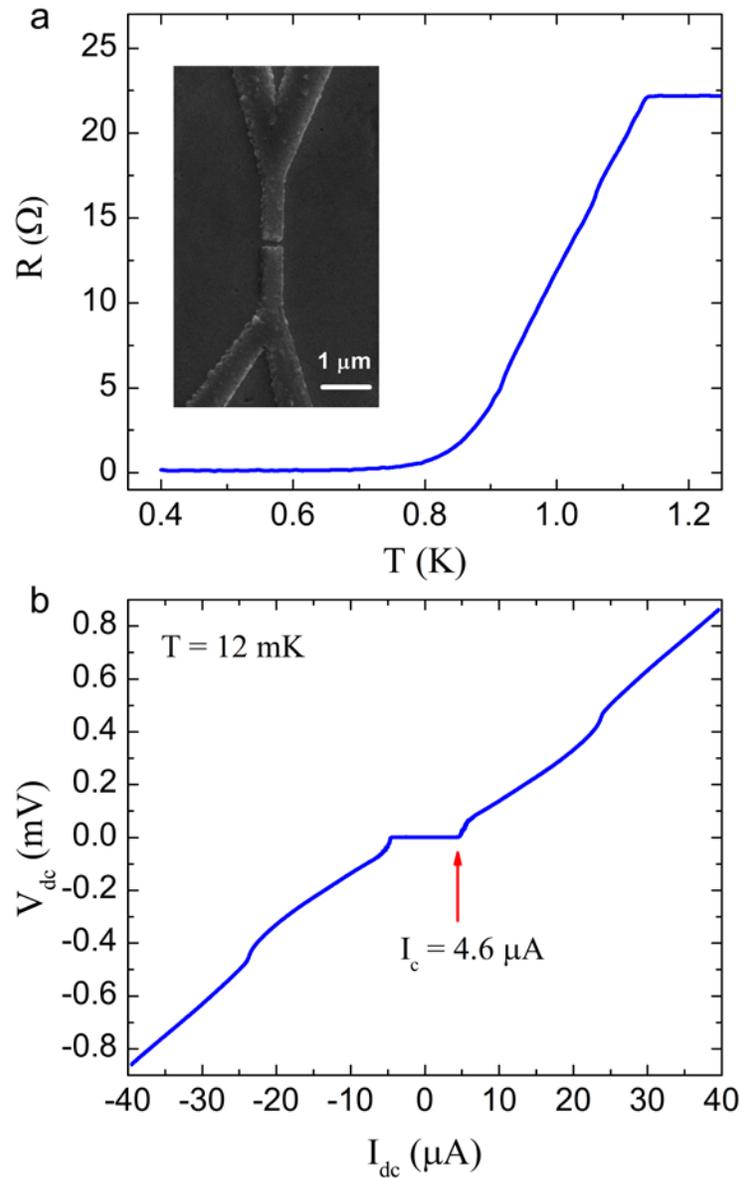

**Figure 1. Experimental configuration and observation of supercurrents in Cd$_3$As$_2$:** (**a**) The resistance of a superconductor-Cd$_3$As$_2$-superconductor junction versus temperature at zero magnetic field. A zero-resistance state is reached below $T_c$ = 0.8 K. Inset shows a scanning electron microscopy image of a Josephson junction. (**b**) The current-voltage (*I-V*) characteristic of Junction A measured at *T* = 12 mK. Supercurrent is observed with a critical current $I_c$ = 4.6 µA.

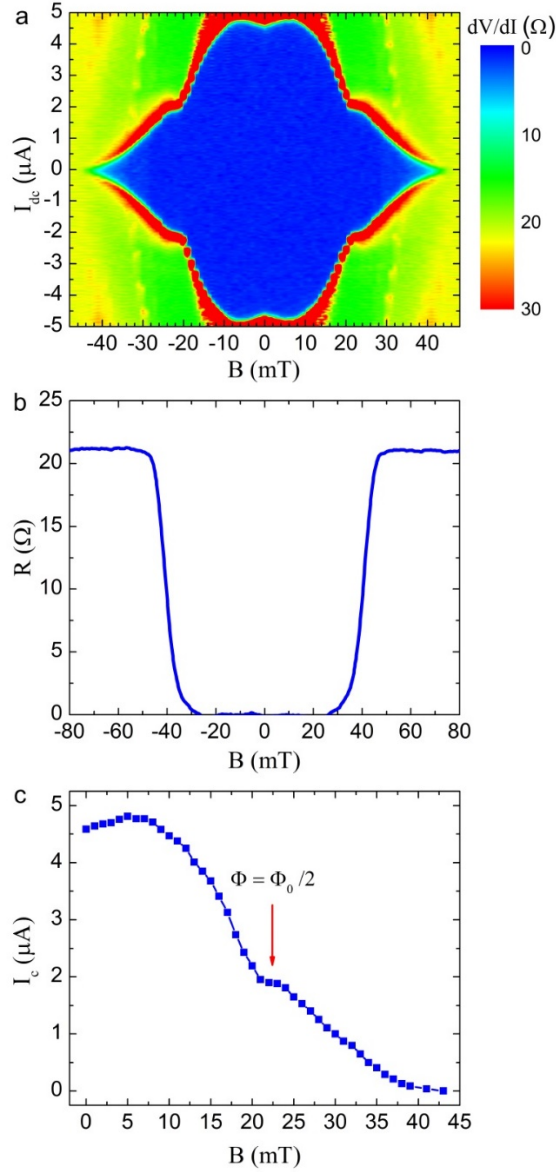

**Figure 2. Current-phase measurements of an Al-Cd$_3$As$_2$-Al Josephson junction:** (**a**) 2D color plot of d$V$/d$I$ with respect to the magnetic field and the d.c. current to better demonstrate the current-phase relationship. The blue area represents the superconducting state and the edge exhibits critical current $I_c$. (**b**) The magneto-resistance of Junction A measured at 12 mK. The resistance remains zero around $B = 0$ and starts to increase at $B = 30$ mT. At $B > 40$ mT, the superconductivity is completely destroyed and the junction is in the normal state. (**c**) $I_c$ in Junction A is plotted as a function of magnetic field. $I_c$ enhancement induced by the magnetic field has been observed in the low magnetic field regime due to the strong spin-orbit coupling. The weak minimum at $B = 22$ mT labeled by the arrow corresponds to $\Phi = \Phi_0/2$ indicating a π-period supercurrent.

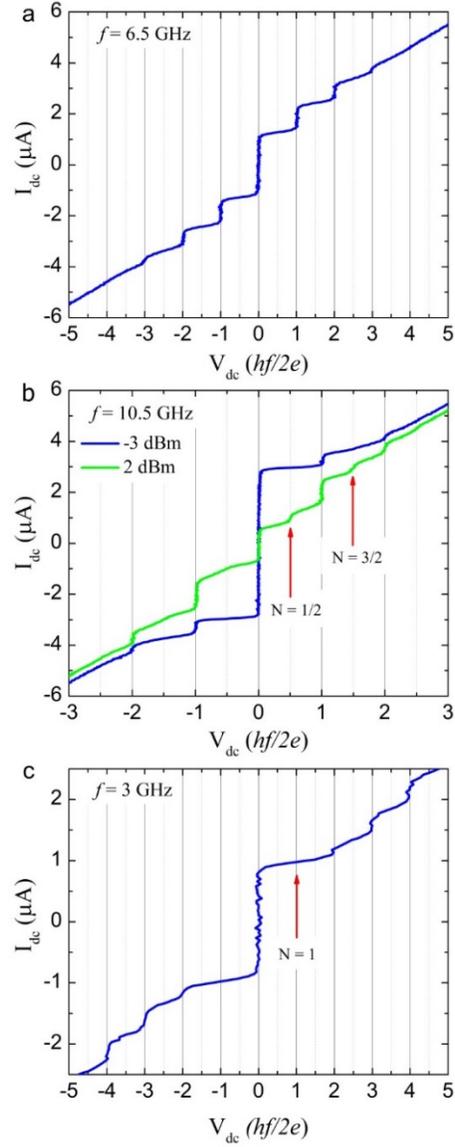

**Figure 3. Observation of the a.c. Josephson effect in an Al-Cd$_3$As$_2$-Al Josephson junction:** (**a**) *I-V* curves measured with microwave irradiation $f$ = 6.5 GHz for different irradiation power values. Integer Shapiro steps are clearly visible at $V_{dc}$ = N$hf/2e$ due to the a.c. Josephson effect. (**b**) *I-V* curves measured with microwave irradiation $f$ = 10.5 GHz. Integer Shapiro steps are observed for low irradiation power $P$ = -3 dBm (blue trace). With increasing irradiation power, fractional Shapiro steps emerge. The green trace shows the *I-V* curve measured with $P$ = 2 dBm. Half integer Shapiro steps are marked by the red arrows. (**c**) *I-V* curves for microwave irradiation of low frequency $f$ = 3 GHz. For high irradiation power, all integer steps are visible. When the power is lowered, step N = 1 is missing indicating a 4π periodic Josephson effect.

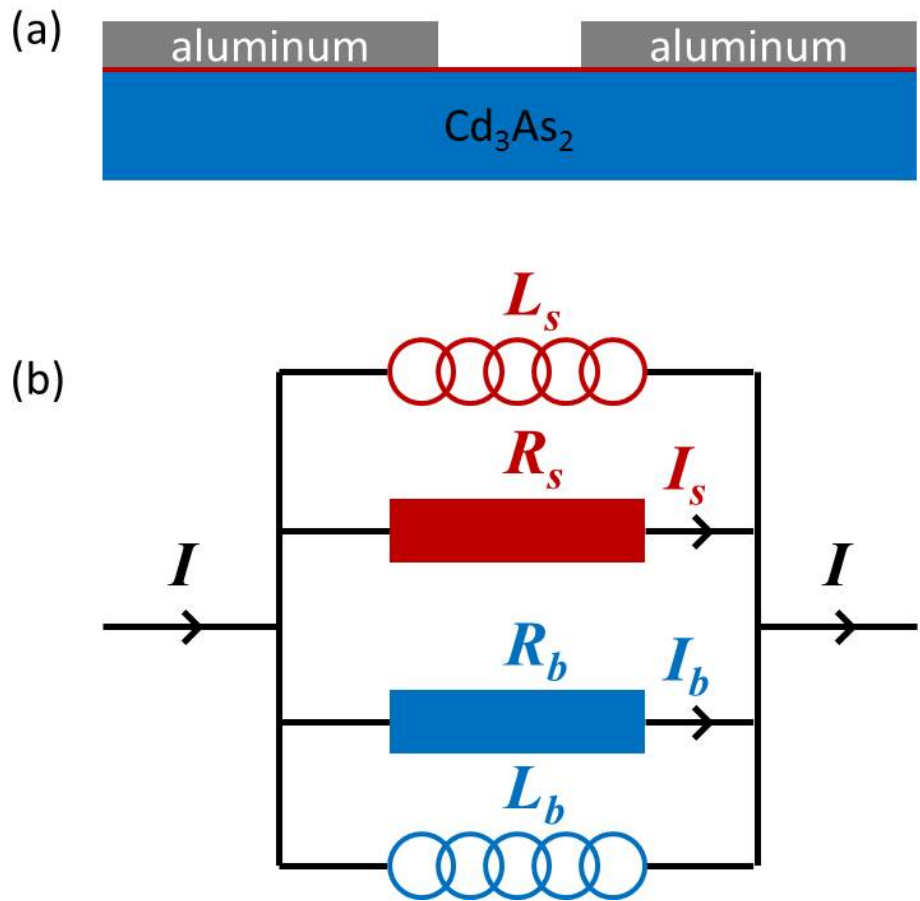

**Figure 4. The SQUID diagram used in our macroscopic model: (a)** Schematic sideview of the superconductor-Cd$_3$As$_2$-superconductor device structure used in our experiment. **(b)** Schematic circuit diagram showing the coexistence of bulk-state (indicated by the blue color) and surface-state (indicated by the red color) channels in the Dirac semimetal. $L_{s(b)}$, $R_{s(b)}$, and $I_{s(b)}$ are the surface (bulk) inductance, resistance, and current.

# SUPPLEMENTAL MATERIAL

## π and 4π Josephson Effects Mediated by a Dirac Semimetal

W. Yu[1], W. Pan[1], D. L. Medlin[2], M. A. Rodriguez[1], S. R. Lee[1], Zhi-qiang Bao[3], and F. Zhang[3]

[1]*Sandia National Laboratories, Albuquerque, New Mexico 87185, USA*

[2]*Sandia National Laboratories, Livermore, CA 94551, USA*

[3]*Department of Physics, University of Texas at Dallas, Dallas, TX, USA*

**x-ray diffraction**

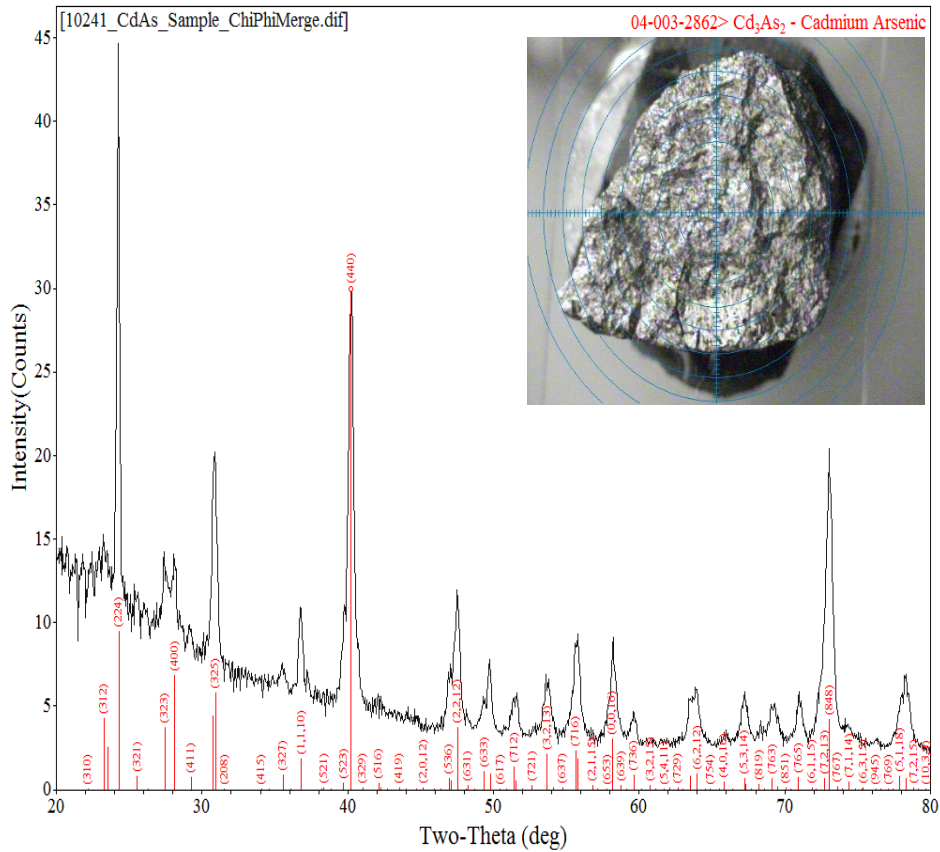

**Figure S1: Equivalent powder-diffraction scan of the $Cd_3As_2$ ingot materials used to fabricate the Al-$Cd_3As_2$-Al Josephson junctions.** All of the observed diffraction maxima correspond with expected reflections for single-phase $Cd_3As_2$ (plotted as vertical red lines along the bottom edge of the figure). The x-ray diffraction (XRD) data were acquired using a Bruker D8 micro-diffraction system equipped with an area detector, with the ingot scanned in two-theta at each of 14 x 60 sample tilt and rotational orientations. The resulting multidimensional diffraction-data sets were post-processed and merged to derive both the equivalent powder-diffraction scan shown in Fig. S1 and the conventional pole-figure representations shown in Fig. S2. The inset shows a photograph of the ~ 8x10x2 mm$^3$ ingot as mounted for the XRD analysis.

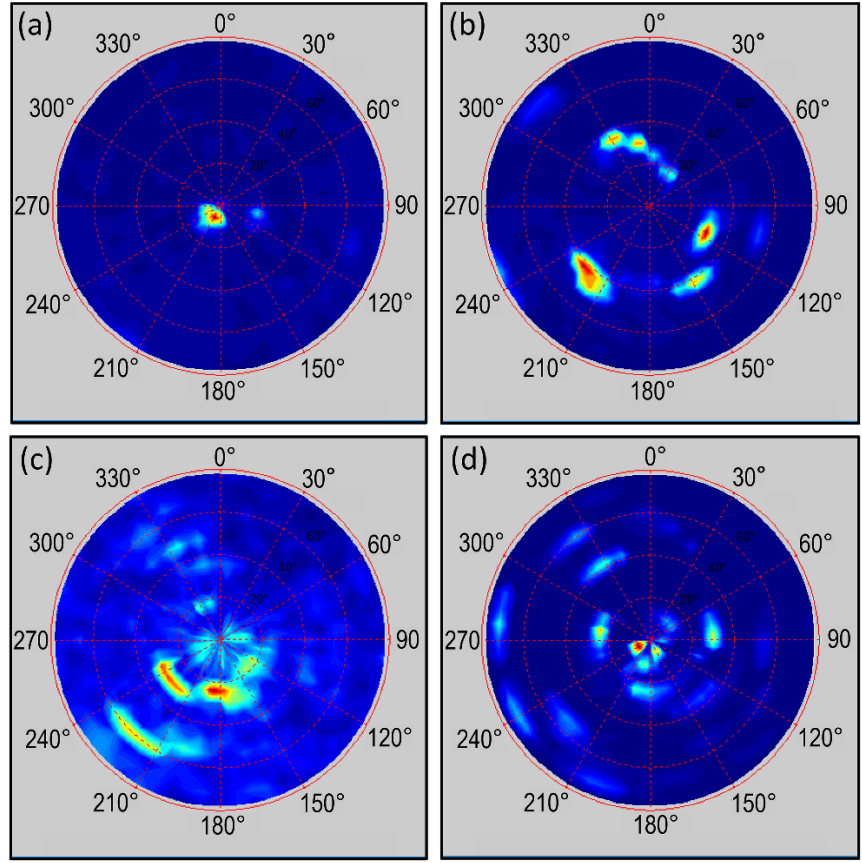

**Figure S2**: **Equivalent XRD pole figures plotted for selected two-theta angular ranges and corresponding Cd$_3$As$_2$ reflections:** **(a)** 2θ = 24.2-24.45° about (224), **(b)** 2θ = 40-40.5° about (440), **(c)** 2θ = 70.7-71.3° about (765), and **(d)** 2θ = 72.9-73.3° about (848). The plotted azimuthal angle indicates the sample rotation (ϕ), and the plotted radial angle indicates the sample tilt (χ), where χ ranges from 0-78°. Relative diffraction intensities are plotted on a normalized linear scale for each pole figure. The observed pole-figure diffraction maxima are relatively small in number and exhibit a discontinuous and spotty character free of any systematic orientation. These diffraction characteristics indicate that the Cd$_3$As$_2$ ingots are comprised of large, randomly oriented polycrystals. The small number of randomly oriented diffracting crystallites explains why the measured peak intensities seen in Fig. S1 do not precisely mirror the relative intensities expected for a random-powder diffraction pattern (shown in red in Fig. S1).

## Cd$_3$As$_2$ thin flakes preparation and magnetoresistance

Cd$_3$As$_2$ thin flakes are obtained from the source ingots by a mechanical method: Two thick, high-purity, polished sapphire wafers are used to capture and crush Cd$_3$As$_2$ ingot fragment to form a fine, granular powder. Then, Cd$_3$As$_2$ thin flakes are simultaneously created and transferred to a Si/SiO2 carrier substrate by sliding the Si/SiO$_2$ substrate across the face of one of the sapphire wafers holding the Cd$_3$As$_2$ powder. In order to make the Cd$_3$As$_2$ flakes as thin as possible, we apply high mechanical pressure to the Si/SiO$_2$ substrate during the final sliding step. Because high mechanical shearing forces are applied to obtain the nm-scale flakes, minute portions of the sliding SiO$_2$ layer may locally fracture to become incorporated into the exfoliated Cd$_3$As$_2$ flakes (see EDS in Fig. S4). Finally, specific flakes are selected for subsequent device fabrication through visual inspection of the carrier substrate by optical microscopy.

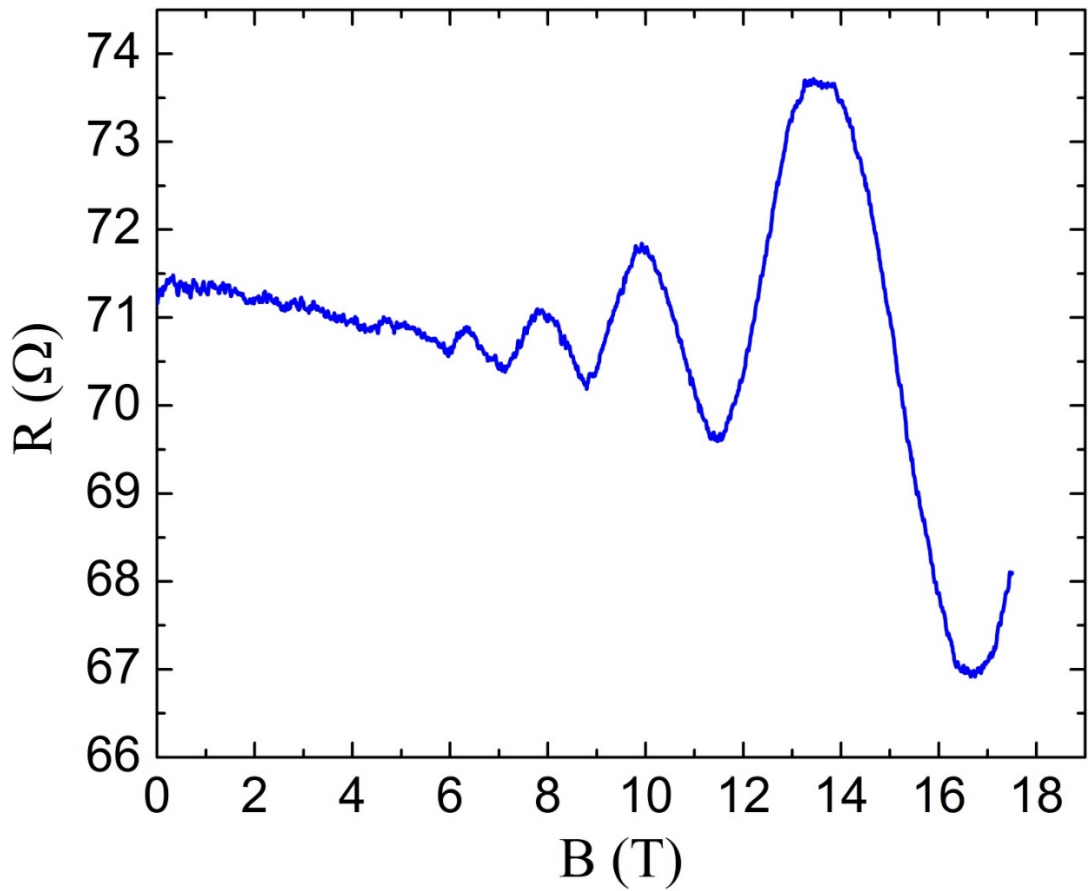

**Figure S3. Quantum oscillations in a thin Cd$_3$As$_2$ flake with nomal contacts.** Magnetoresistance is measured at 0.57 K. Magnetic field is in plane with the Cd$_3$As$_2$ flake. Quantum oscillations are observed indicating Cd$_3$As$_2$ of a high quality suitable for use in transport studies.

**Transmission Electron Microscopy**

We prepared a cross-sectional transmission electron microscopy specimen from a $Cd_3As_2$ device structure using a Focused Ion Beam (FIB) system (FEI Helios Nanolab 660 system, FEI/Thermofisher, Hillsboro, OR). Observations were conducted using an FEI 80-200 Titan Scanning Transmission Electron Microscope (STEM) operated at 200 keV and equipped with a 4-SDD SuperX energy dispersive x-ray spectrometer (EDS). EDS spectra were analyzed using the commercial Bruker Esprit (v 2.1) software package and quantified using library standards provided with the software.

Figure S4(a) presents a bright-field STEM image of the device in cross-section, illustrating the $SiO_2$ of the underlying oxidized Si wafer, the $Cd_3As_2$ flake, and the aluminum contact. An EDS map from the indicated region is shown in Figure S4(b). A higher magnification EDS map is shown in Figure S4(c).

The interfaces between the flake and $SiO_2$ and the Aluminum are sharp and flat. We see no evidence for a significant inter-reaction between the $Cd_3As_2$ and the Aluminum contact. The EDS spectra collected from within the $Cd_3As_2$ layer do detect a weak Aluminum signal that would correspond to less than 2 at. %. However, we believe that this low-level signal is likely an artifact resulting from secondary flourescence of the immediately adjacent Al layer.

We also observe some silicon-rich inclusions in the $Cd_3As_2$ (Figure S4(b)). The origin of these inclusions is unclear. Although they could be present in the original, as grown material, another possibility is that they are incorporated from particles on the wafer as a result of the high shearing forces during the transfer process to the $Si/SiO_2$ wafer.

Considering only the contributions from the Cd and As signals, quantification of the EDS gives a composition of 59±7 at. % Cd and 41 ± 1 at. % As, which is consistent with the expected composition of 60 at. % Cd and 40 at. % As.

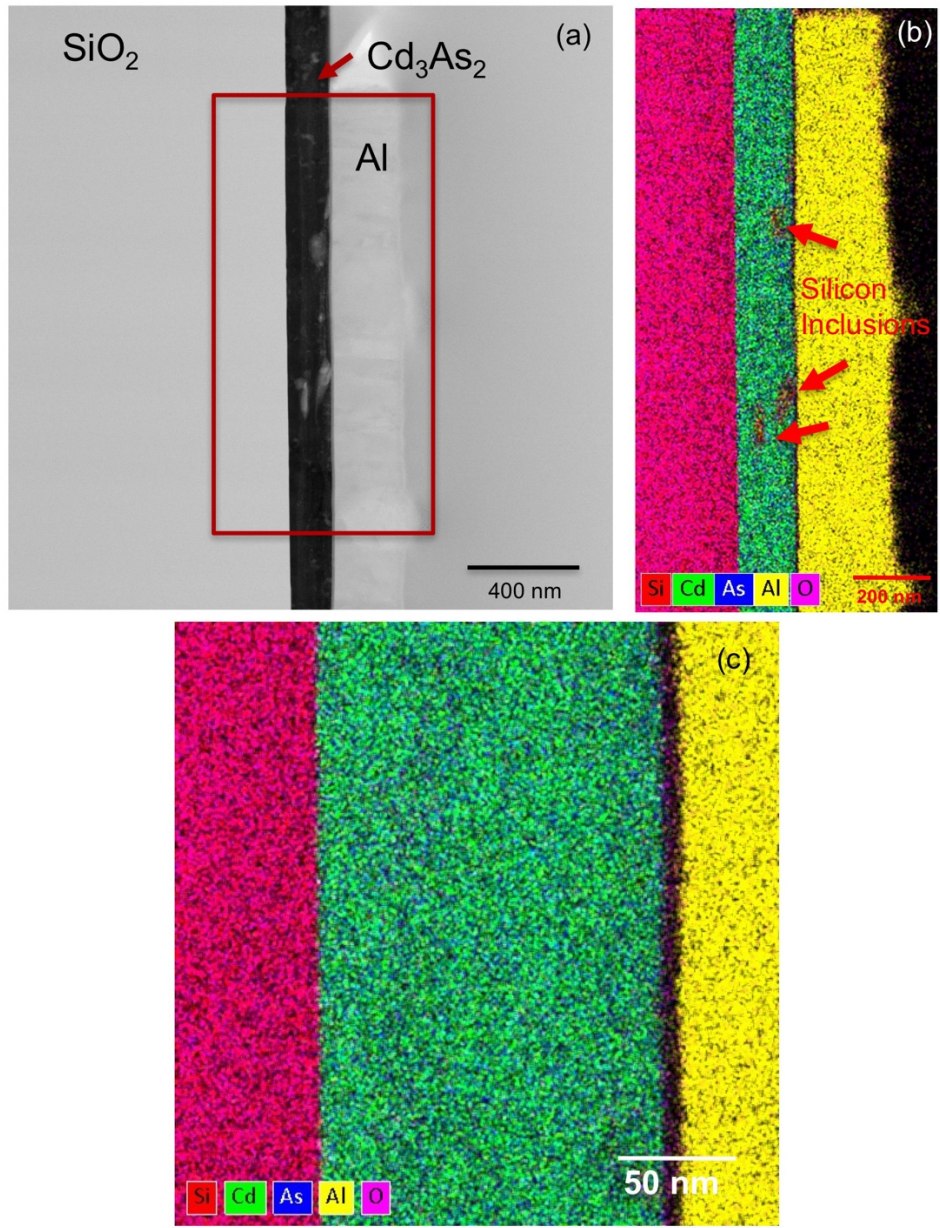

**Figure S4**: **Scanning transmission electron microscopy (STEM) results.** (a) BF-STEM image of a Cd$_3$As$_2$ device structure in cross-section. (b) EDS composite map from the region of (a) marked by the red box. (c) A higher magnification EDS map.

**Junction B**

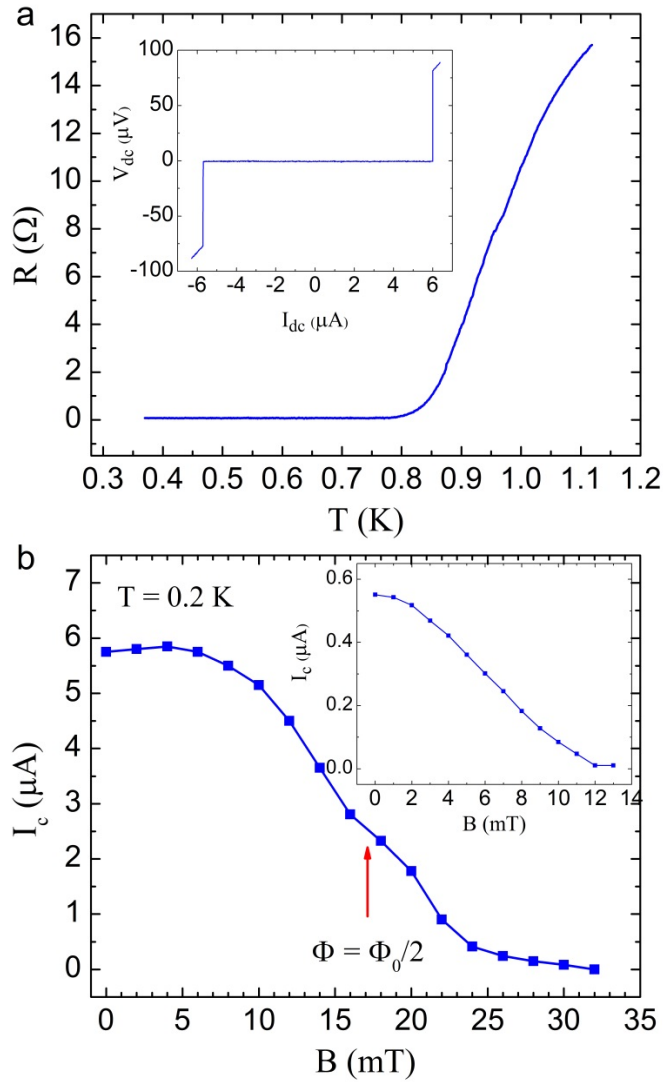

**Figure S5. Transport properties of Junction B: (a)** The resistance of Junction B is plotted as a function of temperature. The critical temperature is about 0.8 K. Inset shows *I-V* curve measured at 0.2K. A clear supercurrent state is observed. **(b)** Critical current $I_c$ is plotted versus *B* at 0.2 K. $I_c$ increases with *B* in the low field regime. A minimum is visible at ~ 17 mT corresponding $\Phi = \Phi_0/2$ indicating $\pi$–period supercurrent. $\Phi_0$ is the magnetic flux quantum. Inset shows $I_c$ versus *B* at 0.7 K. $I_c$ enhancement is suppressed by temperature.

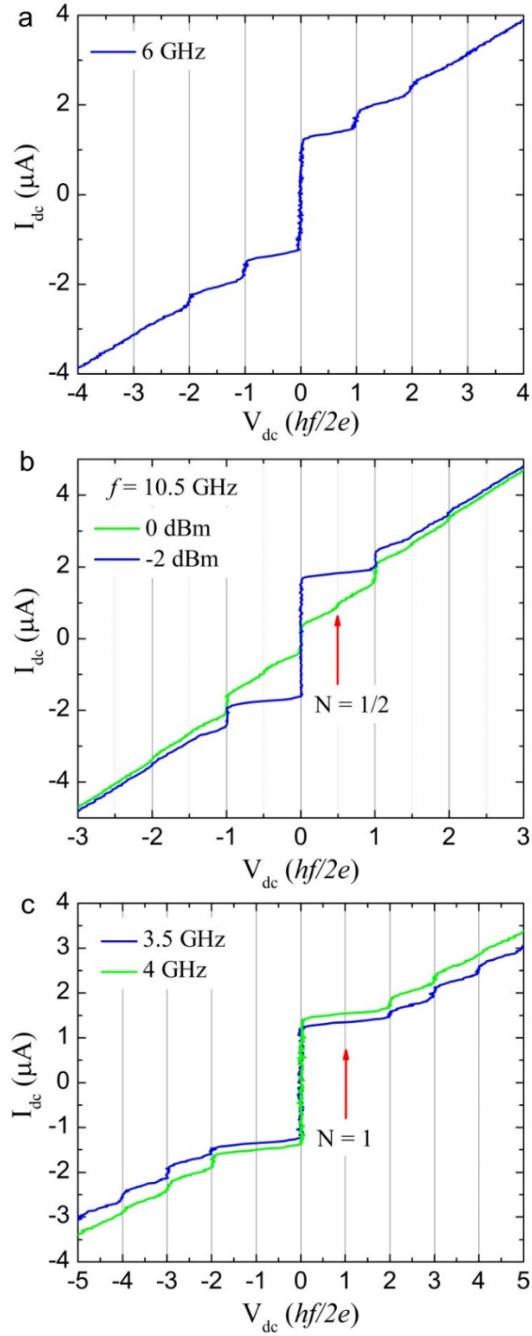

**Figure S6. Shapiro steps in Junction B:** (**a**) *I-V* curves measured with microwave irradiation *f* = 6 GHz Integer Shapiro steps are visible at $V_{dc} = nhf/2e$. (**b**) *I-V* curves measured with microwave irradiation *f* = 10.5 GHz. For low irradiation power, Shapiro steps occur at integer numbers (blue trace). With increasing irradiation power, fractional Shapiro steps are observed as marked by the red arrows (green trace). (**c**) *I-V* curves at different frequencies demonstrating $4\pi$-periodic Josephson effect (Step N = 1 is missing).

# Theoretical Calculations on Anomalous Shapiro Steps

In addition to integer Shapiro steps, half-integer Shapiro steps also emerge in our experiment. In this section, we analyze the origin of both the integer and half-integer Shapiro steps.

Because of the coexistence of surface-state and bulk-state channels in the Dirac semimetal, our Josephson junction can be viewed effectively as a superconducting quantum interference device (SQUID). Such a device can then be described by a resistively shunted junction model [1] as follows:

$$I_j = \frac{V_j}{R_j} + I_{cj} \sin\phi_j,$$
$$\frac{d\phi_j}{dt} = \frac{2e}{\hbar} V_j,$$
$$\phi_b - \phi_s = \frac{2\pi}{\Phi_0}\Phi, \#(1)$$
$$\Phi = L(I_s - I_b) + \Phi_{\text{ex}} + \Phi_{\text{in}},$$
$$V = V_j + L\frac{dI_j}{dt} = V_0 + V_1 \cos\omega_f t,$$

where $j = b$ and $s$ stand for the bulk and surface, respectively. In Eqs. (1), $I_j$, $R_j$, and $I_{cj}$ denote the current, resistance, and superconducting critical current in the $j$-channel; $V_j$ and $\phi_j$ are the voltage and phase difference across the $j$-channel; $\Phi_0$ is the magnetic flux quantum, $L$ is the self-inductance, $\Phi_{\text{in}}$ is the intrinsic phase difference between the bulk and surface channels, $\Phi_{\text{ex}}$ is the phase difference induced by an external magnetic flux if feasible ($\Phi_{\text{ex}} = 0$ in our experiment), and $V$ is the applied voltage.

For simplicity, we choose $R_b = R_s = R$ and $I_{cb} = I_{cs} = I_c$ in our calculation (the conclusion does not change in the more general case) and denote $\tau = 2\pi R I_c t/\Phi_0$, $\beta = LI_c/\Phi_0$, and $i = (I_a + I_b)/I_c$. Thus, Eqs. (1) lead to the following two equations:

$$\frac{d\phi_s}{d\tau} + \sin\phi_s + \frac{\phi_s - \phi_b}{4\pi\beta} = \frac{1}{2}\left[i - \frac{1}{\beta}\left(\frac{\Phi_{\text{ex}}}{\Phi_0} + \frac{\Phi_{\text{in}}}{\Phi_0}\right)\right], \#(2a)$$
$$\frac{d\phi_b}{d\tau} + \sin\phi_b - \frac{\phi_s - \phi_b}{4\pi\beta} = \frac{1}{2}\left[i + \frac{1}{\beta}\left(\frac{\Phi_{\text{ex}}}{\Phi_0} + \frac{\Phi_{\text{in}}}{\Phi_0}\right)\right]. \#(2b)$$

By denoting $\Psi = (\Phi_{\text{ex}} + \Phi_{\text{in}})/\Phi_0$ and expanding Eqs. (2) to the first order in $\beta$ [2], we obtain the current:

$$I = 2\text{Im}\left[xe^{i\phi_0}\sum_{k=-\infty}^{\infty} J_{-k}(a)e^{i(\omega_0 - k\omega)\tau} + \pi\beta y^2 e^{i2\phi_0}\sum_{k=-\infty}^{\infty} J_{-k}(2a)e^{i(2\omega_0 - k\omega)\tau}\right], \#(3)$$

where $x = \cos\pi\Psi$, $y = \sin\pi\Psi$, $\omega_0 = V_0/RI_c$, $a = 2\pi V_1/\omega_f \Phi_0$, $\omega = \Phi_0\omega_f/2\pi RI_c$, $\phi_0$ is an initial constant, and $J_k$ is the $k$-th order Bessel function.

Therefore, the Shapiro steps emerge at $\omega_0 = k\omega$ and $\omega_0 = k\omega/2$ with $k$ being integers, which respectively produce non-oscillating components in the first and second terms of Eq. (3). For appearance of half-integer steps, the second term needs to be non-zero, i.e., the self-inductance $L$ exists and the phase difference between the two channels is not an integer multiple of $2\pi$. Moreover, the half-integer steps become prominent when $L$ is large (but still sufficiently small for the consideration of higher order terms in Eq. (3)) and when the phase difference is close to a half integer multiple of $2\pi$. Given $\Phi_{ex} = 0$ in our experiment, the presence of half-integer Shapiro steps implies that the surface-state and bulk-state channels form a spontaneous 0-$\pi$ junction (at least close to $\pi$).